# Predicting the future success of scientific publications through social network and semantic analysis[1]

Andrea Fronzetti Colladon[a], Ciriaco Andrea D'Angelo[b,2], Peter A. Gloor[c]

[a] *University of Perugia, Department of Engineering*
*Via G. Duranti, 93 - 06125 Perugia, Italy*
andrea.fronzetticolladon@unipg.it

[b] *University of Rome "Tor Vergata", Department of Engineering and Management*
*Via del Politecnico, 1 - 00133 Rome, Italy*
dangelo@dii.uniroma2.it

[c] *MIT Center for Collective Intelligence*
*245 First Street, 02142 Cambridge, MA, USA*
pgloor@mit.edu

**Abstract**
Citations acknowledge the impact a scientific publication has on subsequent work. At the same time, deciding how and when to cite a paper, is also heavily influenced by social factors. In this work, we conduct an empirical analysis based on a dataset of 2010-2012 global publications in chemical engineering. We use social network analysis and text mining to measure publication attributes and understand which variables can better help predicting their future success. Controlling for intrinsic quality of a publication and for the number of authors in the byline, we are able to predict scholarly impact of a paper in terms of citations received six years after publication with almost 80 percent accuracy. Results suggest that, all other things being equal, it is better to co-publish with rotating co-authors and write the papers' abstract using more positive words, and a more complex, thus more informative, language. Publications that result from the collaboration of different social groups also attract more citations.

**Keywords**
*Social network analysis; text mining; social capital; abstract; citability; scholarly impact.*

## 1. Introduction

Measuring the value of a scientific publication is extremely complex but also crucial for many decisions related to research management and science policy. Scientific publications encoding new knowledge have different values, depending on their impact on future scientific advancements and ultimately on social and economic development. As a proxy for such impact, bibliometricians adopt citation-based indicators. The choice of using citation indicators as a proxy for the impact of scientific production is based on assumptions deriving from sociology of science. In a narrative review of studies on the citing behavior of scientists, Bornmann and Daniel (2008), analyze the motivations that push scientists to cite the work of others. The findings show that "citing behavior is not motivated solely by the wish to acknowledge intellectual and cognitive influences of colleague scientists, since the individual studies reveal also other, in part non-scientific, factors that play a part in the decision to cite". Nevertheless, "there is evidence that the different motivations of citers are not so different or randomly given to such an extent that the phenomenon of citation would lose its role as a reliable measure of impact". In particular, previous literature proposes two different theories of citing behavior: the normative theory and the social constructivist view. The first, based on the work of Robert Merton (1957), affirms that scientists, through the citation of a scientific work, recognize a credit towards a colleague whose results they have used. In this case, the citation represents an intellectual or cognitive influence on their scientific work. The social constructivist view on citing behavior is based instead on constructivist theory in sociology of science (Knorr-Cetina 1981; Latour and Woolgar 1986). This approach contests the assumptions at the basis of normative theory and thus weakens the validity of evaluative citation analysis. Constructivists argue that "scientific knowledge is socially constructed through the manipulation of political and financial resources and the use of rhetorical devices" (Knorr-Cetina 1991), with the direct consequence that citations are not linked in direct and consequential manner to the scientific contents of the cited article. The bibliometric approach is based instead on the assumption that this link is strong and direct, meaning that citational analysis can be the principal instrument for evaluating the impact of scientific production.

We agree with this assumption and the Mertonian, normative concept of what citations signify, although there might be exceptions (uncitedness, negative citations, fraudulent cross-citations, etc.). Although both theories have their merits, it still remains to understand: i) which of them better explains the citability of a scientific work and; ii) whether there are other determinants of citability not covered by these two theories. On this last issue, we assume that there are "hidden honest signals" underlying the cognitive and intellectual process that produces a paper, that draw the attention of readers, influencing its citability, beyond its intrinsic quality and the social capital of its authors. Before "citing" a paper, a scholar needs to read it. Therefore maybe some semantic features related to the content of a paper (and, consequently, to its cognitive/intellectual appeal) might explain its readability and accessibility, and therefore its subsequent citability. When analyzing literature on a given topic, scientists generally rely on websites of journals' publishers, on bibliometric platforms (WoS, Scopus), or on science social media (Mendeley, Academia, Researchgate, Google Scholar). Before downloading and reading the full text, they analyze the abstracts resulting from a specific search query. We wonder if some features of a publication's abstract might affect its readability and, therefore, its citability. It is known that it matters "what" you publish and "with whom": now, we want to investigate whether the "how" also counts, meaning "how an author sells" (in the abstract) the outcomes of her/his research to prospective readers and, as a consequence, to prospective citers.

In this work, we propose an empirical analysis based on a dataset of 2010-2012 worldwide publications in chemical engineering, indexed in SCOPUS. In particular, we compare publication metrics at the time of publishing, with scholarly impact of the paper six years after publication. Controlling for intrinsic quality of a publication, proxied by the impact factor of the journal and by the number of authors in the byline, we aim at understanding the importance of three sets of predictive variables: structural social network metrics, dynamic changes in network position of authors, and complexity and sentiment of abstracts. The first two sets of variables complement the cardinality of the byline, in proxying the "social capital" of authors. The third set tries to catch cognitive/intellectual appeal of a paper based on semantic features of its summary offered to the reader.

Using machine learning, we are able to predict the impact of a paper in terms of numbers of citations six years after publication with 79 percent accuracy. We found that it is better to co-publish with many well-connected authors and write the abstract using more positive words, and employing more complex, thus more informative, language.

The next section offers a picture of previous literature on different issues related to our analysis; section 3 illustrates methodological issues of the work, i.e. data collection and variables of the inferential model; section 4 presents results of the analysis; section 5 closes the work discussing results and proposing concluding remarks.

## 2. Literature review

Our paper aims at analyzing the predictability of long-term citations received by a publication observing the social capital of the authors in the byline and the features of its abstract. A summary of the main contributions of these two literature streams will be presented below.

### 2.1. Social capital, research collaboration and impact of co-authored publications

The scientific environment is no different from other human activities by requiring to work in cooperation, because the individual scientist cannot possess all the competencies and resources needed for the resolution of the problem she/he is working on. Three concomitant factors help explain the remarkable increase of collaboration among scientists, research groups, and institutions, witnessed during the last decades: i) the increasing complexity and cost of research to solve global societal problems, mostly interdisciplinary in nature (Bennett and Gadlin 2012; Persson et al. 1997); ii) the general reduction in travel costs, as well as the diffusion of inexpensive new communication technologies, in particular the Internet, which has greatly reduced the qualitative divide between distant and face-to-face communication (Hoekman et al. 2010; Olson and Olson 2000); iii) the existence of incentive systems towards collaborative research (Defazio et al. 2009). These factors have a systemic impact: at the level of individuals, they encourage scientists to increase their own "social capital", defined as the whole of the resources obtainable through one's social network (Jha and Welch 2010). Such resources include both the social network itself and those that are accessible via the network. For Nahapiet and Ghoshal (1998), social capital is a concept involving three dimensions: structural, cognitive and relational. The structural dimension concerns the general degree of connection and density of the network structure. The cognitive dimension concerns the sharing of knowledge between the actors of the network; the relational dimension concerns the quality of interpersonal relations in terms of trust, respect, friendship, etc. The relational dimension is the one that most influences the availability and use of resources in a social network of researchers (Burt 1995).

In the context of research systems, social capital is integral to the more encompassing concept of scientific and technical human capital (S&T human capital; STHC). Social capital and STHC are highly interdependent. Each enables growth of the other. To be able to grow their social capital, scientists have to develop some basis of STHC over the course of their early career, in order to catch the interest of other colleagues (Bozeman et al. 2001; Dietz 2000; Murray 2005). It is no accident that scientists with tenure and the largest research projects tend to have larger, more heterogeneous and cosmopolitan, collaboration networks. They expand their networks beyond home institutions (Bozeman et al. 2001; Bozeman and Corley 2004) and countries (Melkers and Kiopa 2010). As social capital increases, the potential intensity and quality of research collaboration increases in parallel with growth in STHC. Scientists use their social networks for multiple purposes, including the identification and selection of collaborators (Beaver 2004; Katz and Martin 1997; Maglaughlin and Sonnenwald 2005). According to Wagner, Park and Leydesdorff (2015), future stars consciously build collaboration networks with other future stars well before they become famous. Sekara et al. (2018) have identified a "chaperone effect" where senior highly cited researchers help junior researchers in their team to establish themselves in a field and acquire senior status themselves. On the other hand, analyzing the scientific impact of a platform's programming community that produces digital scientific innovations, Brunswicker, Mate, Zentne, Zentner, and Klimeck (2017) state that being surrounded by star performers can be harmful.

The impulse to undertake research collaboration studies has been supported by the development of specific bibliometric tools, which permit measurement of the different dimensions that characterize the phenomenon. In the literature, bibliometrics and the analysis of co-authorships have become the standard ways of observing research collaborations and measuring social capital. It should also be noted that co-authorships should be handled with care as a source of evidence for true scientific collaboration: this assumption has been questioned by many bibliometricians (Kim and Diesner 2015; Laudel 2002; J. Lundberg et al. 2006; Melin and Persson 1996). As Katz and Martin (1997) stated, some forms of collaboration do not generate co-authored articles and some co-authored articles do not reflect actual collaboration. However, in contradiction to the limitations noted above, this approach offers notable advantages both in terms of sample sizes (and consequent power of analysis) and of cost-effectiveness.

Social Network Analysis (SNA) is frequently used in the evaluation of the scientists' social capital. The diffusion of collaboration studies based on SNA was particularly stimulated by Melin and Persson (1996), whose seminal study outlined procedures for the construction and analysis of co-authorship networks. In the literature on research collaboration, indicators of centrality have often been used in attempts to validate hypotheses related to social capital theory (Jha and Welch 2010; Nahapiet and Ghoshal 1998) and the contextual development of human and social capital (Bozeman et al. 2001; Bozeman and Corley 2004). A subject of great attention has been the so-called mechanisms of preferential attachment, meaning that when a scientist begins publishing, she will tend to collaborate with other scientists having a higher level of degree centrality (Barabási et al. 2002; M. Li et al. 2007; Perc 2010). In this manner, the cumulative advantage of the most popular scientists increases, in line with the *Matthew Effect* (R K Merton 1968), and the role of the hub within the network continues to strengthen. Previous research suggested the existence of a tight relationship between the number of authors in the byline and the long-term citation impact of publications (Abramo and D'Angelo 2015; Bornmann et al. 2014; Franceschet and Costantini 2010; Larivière et al. 2015; Matveeva and Poldin 2016; Waltman and van Eck 2015).

A few studies have focused on how centrality indicators of authors interact and affect citations for publications. In general, these studies claim that a papers' citations are related to the node attributes of their authors in the collaboration network. The only exception was presented by Wang (2014) who, exploring the Matthew effect, found no impact of authors' networking and prestige on solo-authored papers' citations. By contrast, working on a sample of more than 30 thousand authors in Google Scholar, Matveeva and Poldin (2016) discovered a positive relationship between scholars' citation counts and authors' centrality. Using wind-energy paper data collected from WoS, Guan, Yan, and Zhang (2017) found that the structural holes of authors have positive but non-significant effects on a paper's citations, while the authors' centrality has an inverted U effect.

Lastly, Li, Liao, and Yen (2013) defined six specific indicators of co-authorship network characteristics according to the social capital theory and provided several strategies for leveraging social capital, meant to support scholars who want to enhance their research impact.

As better detailed in Section 3.1, for measuring the "social capital" of authors, along with the cardinality of the byline, we propose both structural social network metrics and the analysis of dynamic changes in network position of authors. To the best of our knowledge, this last set of variables represents a novelty compared to previous studies on the same topic.

## 2.2. The influence of textual content on the citability of publications

Many authors have investigated the impact of factors other than intrinsic quality and authors' social capital on publication citations. Bornmann, Leydesdorff, and Wang (2014) showed that the number of cited references, and the number of pages are useful covariates in the prediction of long-term citation impact. Others have tested the effect of the presence of a country's name in the title (Abramo et al. 2016; Jacques and Sebire 2010; Nair and Gibbert 2016; Paiva et al. 2012) or of the ordering of authors in the byline (Abramo and D'Angelo 2017; Huang 2015; Ong et al. 2018; Shevlin and Davies 1997). Other studies have concentrated on the importance of the article title because, as Haggan (2004, p. 293) reasons, "the title plays an important role as the first point of contact between writer and potential reader and may decide whether or not the paper is read". We point out a set of works on the relation between the structure of the title and citation rates (Habibzadeh and Yadollahie 2010; Jacques and Sebire 2010; Jamali and Nikzad 2011; Subotic and Mukherjee 2014). Falahati, Goltaji and Parto (2015) conducted a morphological analysis of titles, to study the link between citability and title length/number of punctuation marks. The results of the analysis, made on a sample of 650 articles published in the journal Scientometrics over the years 2009-2011, show that: i) title length and article citations are not correlated; ii) the number of punctuation marks does not serve as a reliable predictor of citations. Habibzadeh and Yadollahie (2010) studied the correlation between the length of an article title and the number of citations, for the area of the medical sciences. Longer titles seem to be associated with higher citation rates, with a larger effect for articles published in journals with a high impact factor. Using a sample including all the articles published in six PLOS journals, Jamali and Nikzad (2011) investigated the influence of the type of article title on the number of citations and downloads that an article receives. They observed that: i) "question" articles tend to be downloaded more often, but cited less compared to others; ii) articles with longer titles are downloaded less than those with shorter titles; iii) titles with colons tend to be longer, and therefore receive less downloads and citations. Rostami, Mohammadpoorasl and Hajizadeh (2014) studied the association between some features of titles relative to the number of citations, examining the articles of the 2007 volume of Addictive Behavior: their results indicate that the type of title, as well as the number of keywords different from the words in the title, can contribute to predicting the number of citations. Uddin and Khan (2016) showed that author selected keywords have a positive impact on the long-term citation count. van Wesel, Wyatt, and ten Haaf (2014) focused their attention on what they call "superficial factors" influencing citations, including the number of words in title, number of pages, number of references, but also sentences in the abstract and readability in general. In fact, if the title plays an important role as a "touch point" for attracting the reader towards the manuscript, the abstract should do so even more by "advertising" its content and encouraging the full reading of the paper. According to Plavén-Sigray, Matheson, Schiffler, and Thompson (2017), the abstracts reflect the overall writing style of entire articles and "the readability of scientific texts is decreasing over time" and this should worry scientists and the wider public, as they impact both the reproducibility and accessibility of research findings. As for the the influence of the abstract on the citability of a publication, Weinberger, Evans, and Allesina (2015) found that shorter abstracts (fewer words and fewer sentences) consistently lead to fewer citations, with short sentences being beneficial only in Mathematics and Physics. Similarly, using more (rather than fewer) adjectives and adverbs is beneficial. Different conclusions are reached by Letchford, Preis, and Moat (2016) who found that journals publishing papers with shorter abstracts and containing more frequently used words receive on average slightly more citations per paper. Lastly, Freeling, Doubleday, and Connell (2019) suggested that increases in clarity, narrative structure, and creativity in the abstract of a paper could translate to a boost in citations it receives.

As better detailed in Section 3.1, in order to assess the possible dependence of citations accrued by a publication, by the cognitive/intellectual appeal of its content, we consider semantic features of the abstract and, specifically, its length, sentiment, complexity, diversity, and commonness. In terms of sentiment, our approach is partially explorative, as only few studies addressed the topic of extraction of opinions from scientific literature so far. In general, we would expect an objective, factual-based communication style used in scientific abstracts – i.e. a more technical language than the one appearing on news, reviews or narrative texts (Athar 2011; Justeson and Katz 1995). However, some studies showed that technical terms can convey sentiment as well, and that "sentiment carrying science-specific terms exist and are relatively frequent" (Athar 2011 p.82; Athar and Teufel 2012; Athar 2014).

## 3. Data Collection and Methodology

Our dataset is made of publications indexed in Scopus in 2010-2012 and hosted by sources tagged as "Chemical engineering" with respect to the ASJC (All Science Journal Classification) schema[3]. The choice of Scopus as bibliometric source is due to a powerful feature available on this repository, the author name disambiguation system[4]: for each publication SCOPUS provides not only the authors' list but also a list of unique codes associated with each author. Kawashima and Tomizawa (2015) estimated the accuracy of the author identification in Scopus and found a recall and precision for Japanese researchers of about 98% and 99% respectively, which makes us particularly confident in terms of accuracy of the social networks that we will analyze.

The choice of the three-year time window maximizes the tradeoff between computational effort and the robustness of the analysis (Wallace et al. 2012); in fact, scientific production is subject to uncertainty due to: i) personal events, ii) patterns in research projects; iii) editorial and indexing processes (Luwel and Moed 1998; Trivedi 1993), iv) accidental facts and errors in bibliometric repositories (Karlsson et al. 2015). According to Abramo, D'Angelo, and Cicero (2012) a three year publication period is appropriate for filtering randomness and assessing research performance and collaboration.

The focus on a specific field poses on the one hand problems of possible generalizability of results, on the other hand is necessary for a smaller-scale analysis as we are doing here, because all the variables at stake are field specific: the intensity of publication and citation, collaboration patterns, structural features of social networks, etc.

---

[3] See https://service.elsevier.com/app/answers/detail/a_id/15181/supporthub/scopus/ for details. Last accessed on March 19, 2020.
[4] https://www.scopus.com/freelookup/form/author.uri. Last accessed on March 19, 2020.

For the construction of the dataset we directly queried SCOPUS through the advanced search box, which returned almost 298,000 records. Given the aim of our analysis, it was necessary to eliminate about 74,000 of these results lacking impact metrics of the hosting source or abstracts. We focused in our analysis on research articles published on scientific journals – excluding reviews, conference papers, book chapters and other document types, such as letters, which appeared much less frequently. The final dataset was made of 223,558 publications, indexed in 657 unique sources. For each publication in the 2012 dataset we counted citations on January 1st, 2019, meaning that the citation window is 6 years. If we exclude the so-called "sleeping beauties", a term coined by van Raan (2004) for indicating papers whose importance is not recognized for several years after publication, this is an adequate citation window for predicting long term impact of publications (Abramo et al. 2011), especially in chemical engineering, a subject category characterized by significant "immediacy", i.e. high speed in reaching the peak of citations. As for the impact of the hosting source we use the Scimago Journal Ranking-SJR, 2012 edition (Guerrero-Bote and Moya-Anegón 2012).

As shown in Table 1, in this period, we register an increase in both the average number of co-authors per publication (from 4.22 in 2010 to 4.49 in 2012) and the share of "collaborative" publications (the share of solo-author papers drops from 6.8% in 2010 to 4.5% in 2012). These figures are fully in line with previous literature indicating a worldwide increase in scientific collaborations (Milojevi 2014), attested both by a rapid decline of the share of single-authored publications (Uddin et al. 2012), and by a significant increase in the average number of authors per publication (Larivière et al. 2015).

*Table 1: Bibliometric dataset*

|  | Year | 2010 | 2011 | 2012 | Total |
|---|---|---|---|---|---|
|  | Unique authors | 199497 | 224462 | 241205 | 498598 |
|  | Publications | 68599 | 76514 | 78445 | 223558 |
|  | Solo author paper | 6.8% | 5.9% | 4.5% | 5.7% |
| No. of authors | Average | 4.22 | 4.33 | 4.49 | 4.35 |
|  | Max | 202 | 125 | 37 | 202 |
|  | St. Dev. | 2.39 | 2.35 | 2.30 | 2.35 |
| Cites | Average | 32.6 | 29.5 | 26.1 | 29.3 |
|  | Max | 5815 | 6759 | 6126 | 6759 |
|  | St. Dev. | 78.7 | 67.4 | 55.5 | 67.4 |

### 3.1. Study Variables

As described in the previous sections, our intent is to evaluate the importance of authors' social capital and semantic structure of abstracts, in predicting scientific success of papers, measured in terms of citations received six years after publication.

In doing so, we must control for the number of authors in the byline and for the impact factor of the hosting source. Journal impact metrics are generally aggregated measures of the impact of hosted articles: high impact articles are published in high impact journals and viceversa (Leimu and Koricheva 2005; Mingers and Xu 2010). Of course, there are evident exceptions and bibliometricians suggest not to use impact factors for measuring the quality and impact of individual publications (Marx and Bornmann 2013; Moed and van Leeuwen 1996; Petersen et al. 2019; Weingart 2005). However, here we must control for the intrinsic quality of a paper without having any other information available than the impact of the hosting journal (in our case the SJR).

As for the social capital of authors, the publication data retrieved from Scopus allowed us the construction of two social networks: the first, which we call *author network*, linking authors who collaborated in the writing of one or more papers; the second, which we call *publication network*, linking publications which share one or more authors. Both networks correspond to undirected graphs, where we indicate with *n* the number of nodes and *m* the total number of edges. In the author network, nodes represent scholars and there is an edge between two nodes if the corresponding scholars wrote at least one paper together; edges are weighted according to the number of co-authored papers. We use this network to evaluate the social capital of authors and their co-publication patterns. In the publication network, on the other hand, nodes represent publications, connected by edges weighted by the number of authors they share. Therefore, if paper A shares three authors with paper B, there will be a link connecting nodes A and B of weight equal to three. This second network tracks the social position of a publication, given the relationships maintained by its authors. Considering the above-mentioned graphs, we were able to calculate well-known centrality metrics, in order to study the network position of each publication and of its authors.

*Degree Centrality*. It corresponds to the number of direct links of a network node, weighted by summing the weights of its adjacent arcs (Freeman 1979; Wasserman and Faust 1994). In the author network, it represents the total strength of the direct connections a node has. In the publication network, it counts how many times the authors of a paper are shared with other papers in the network.

*Betweenness Centrality*. This very well-known centrality metric measures how many times a node lies in-between the shortest network paths that connect the other nodes (Freeman 1979; Wasserman and Faust 1994). Nodes with high betweenness centrality often serve as indirect connection between other pairs of nodes, thus having high brokerage power (Borgatti et al. 2013). Betweenness of node *I* can be calculated according to the following formula (Wasserman and Faust 1994):

$$B(i) = \sum_{j<k} \frac{g_{jk}(i)}{g_{jk}}$$

where $g_{jk}$ is the number of shortest network paths linking the generic pair of nodes *j* and *k*, and $g_{jk}(i)$ is the number of that paths that include node *i*. The formula can be normalized dividing it by its maximum $(n-1)(n-2)/2$.

*Closeness Centrality*. It measures the embeddedness of a node in the social network. The higher the closeness of a node, the shorter the network paths that connect it to its peers. To put it in other words, closeness is measured as the reciprocal of the sum of the length of the shortest paths between the node and all other nodes in the graph (Freeman 1979; Wasserman and Faust 1994):

$$C(i) = \frac{1}{\sum_{j=1}^{n} d_{ij}}$$

where $d_{ij}$ is the length of the shortest path connecting nodes *i* and *j*. Closeness can be normalized, multiplying its value by $(n-1)$, which is its maximum and reflects the case of node *i* being adjacent to all other nodes.

*Constraint (Structural Holes)*. It measures the value of network constraint, for each node (either author or publication), as presented in the work of Burt (1995). The idea behind this metric is that nodes which can mediate across unconnected peers are less constrained by their ego-network, thus also having higher social capital (Burt 2004). For instance consider an example with three nodes, A, B and C, where A is linked to B and C, but a link between these last two is missing. That missing link is called "structural hole" and gives social advantage to A that could mediate interactions between B and C, thus being less "constrained" by its ego-network. This is something A could not do, if B and C were directly connected.

*Rotating Leadership*. It counts the number of oscillations in betweenness centrality an author has in the network, considering subsequent publication years, i.e. if the author's betweenness centrality changes significantly from one year to the other, reaching local maxima or minima (Allen et al. 2016; Kidane and Gloor 2007). Rotating leaders are authors which frequently change their network position, not remaining statically central or peripheral. This metric largely proved its potential in past research, which showed, for example, that rotating styles can favor both online community growth (Antonacci et al. 2017) and startups' innovative performance (Allen et al. 2016).

The first four SNA metrics are calculated for both the *author network* and the *publication network*. The *Rotating Leadership* relates to the *author network* only, so that we have a total of nine metrics.

Analyzing the abstract of each publication, we derived metrics of text mining and semantic analysis, to see which variables related to publication content affect its future scholarly impact. Prior to the calculation of these metrics, we processed abstracts in order to remove those words which give little contribution to the text, such as the words "the" or "and", also known as stop-words. Moreover, we removed word affixes to reduce each word to its stem – a procedure known as stemming, which was carried out using the NLTK package and the Python programming language (Perkins 2014). After this preprocessing phase, we proceeded in calculating:

*Abstract Length*, i.e. the number of text characters in the abstract.

*Sentiment*. It measures the positivity or negativity of the language used in a paper abstract, by means of the VADER rule based model for sentiment analysis (Hutto and Gilbert 2014), included in the NLTK python package. Values range from -1 to 1, where positive values represent a positive average sentiment and negative values correspond to the expression of negative feelings. Even if not context-specific, the VADER lexicon showed a good performance in past research (e.g., Hutto and Gilbert 2014; Newman and Joyner 2018).

*Complexity*. Lexical complexity of an abstract is measured by looking at the standard deviation of the frequency distribution of words used in the text. This metric – successfully used in past research (e.g., Fronzetti Colladon and Vagaggini 2017; Gloor, Fronzetti Colladon, Giacomelli, et al. 2017; Gloor, Fronzetti Colladon, Grippa, et al. 2017) – originates from the idea that there is a number of common words which will occur more often in a text, but when more complex ideas are presented different words will appear, thus increasing the variance of the word frequency distribution. Higher scores indicate higher complexity.

*Lexical Diversity*. Is measured as the ratio of different unique word stems to the total number of words used in an abstract (Malvern et al. 2004).

*Commonness*. This metrics examines the uniqueness of words used in each abstract, based on their overall frequency in all text documents. In a first step, the overall frequency of each word is computed (excluding stop-words and after stemming), considering all abstracts. Subsequently, frequencies are averaged for all words of a single abstract, to assess its commonness. If words used are common to all other abstracts then commonness will be high. Conversely, distinctive abstracts will use words that appear less frequently.

We also tested other variants for complexity, lexical diversity and commonness metrics. One approach was to measure complexity as the likelihood distribution of words within an abstract, i.e. the probability of each word to appear in the text based on the term frequency/inverse document frequency (TF-IDF) information retrieval metric (Brönnimann 2014). However, different metrics did not lead to better results.

In the end we have:

- two control variables: the SJR of the hosting journal and the number of co-authors of the publication;
- nine variables related to social capital of its authors, i.e. their social network position and oscillations ($X_1$-$X_9$), and;
- five variables related to article content, measured by the semantic analysis of its abstract ($X_{10}$-$X_{14}$).

Table 2 shows main descriptive statistics for all the above variables. Note that networks were built considering all publications in the dataset (2010-2012). To properly assess authors' collaboration patterns but in order not to use future information, predictions were carried out only for 2012 publications[5], excluding those with incomplete data (for the byline, abstract, citation count, or SJR).

*Table 2: Descriptive statistics of variables used in the analysis*

| Variable Code | Variable Name | Unit of analysis | M | SD |
| --- | --- | --- | --- | --- |
| Y | Citations | Publications 2012 | 20.65 | 32.092 |
| SJR | SJR | Publications 2012 | 1.630 | 1.261 |
| No.authors | Number of Authors | Publications 2012 | 4.400 | 2.184 |
| $X_1$ | Degree - publication network | Publications 2010-2012 | 18.760 | 24.091 |
| $X_2$ | Constraint - publication network | Publications 2010-2012 | 0.395 | 0.298 |
| $X_3$ | Closeness - publication network | Publications 2010-2012 | 0.233 | 0.291 |
| $X_4$ | Betweenness - publication network | Publications 2010-2012 | $1.417 \times 10^{-5}$ | $5.050 \times 10^{-5}$ |
| $X_5$ | Degree - author network | Publications 2010-2012 | 19.619 | 21.627 |
| $X_6$ | Constraint - author network | Publications 2010-2012 | 0.483 | 0.238 |
| $X_7$ | Closeness - author network | Publications 2010-2012 | 0.291 | 0.321 |
| $X_8$ | Betweenness - author network | Publications 2010-2012 | $7.393 \times 10^{-5}$ | $2.652 \times 10^{-4}$ |
| $X_9$ | Rotating Leadership | Publications 2010-2012 | 2.130 | 2.083 |
| $X_{10}$ | Abstract Length | Publications 2012 | 1135.950 | 431.963 |
| $X_{11}$ | Sentiment | Publications 2012 | 0.516 | 0.525 |
| $X_{12}$ | Complexity | Publications 2012 | 0.869 | 0.395 |
| $X_{13}$ | Diversity | Publications 2012 | 0.745 | 0.099 |
| $X_{14}$ | Commonness | Publications 2012 | 23567.090 | 5610.448 |

## 4. Results

Table 3 shows correlations of the variables at stake. Since they are often not normally distributed and the relationships among them were not necessarily linear, we used a nonparametric approach, i.e. the Spearman's rank correlation (Spearman 1904).

---

[5] - This prevent the need for normalizing citation count, since all publication used for prediction are of the same year and subject field.

As the table shows, many of our predictors significantly and positively correlate with the number of citations accrued by publications after six years. Journal ranking is the one with the strongest correlation. In addition, the number of authors and their position in the author network seem to play an important role: citations are higher for those papers whose authors are more central in terms of direct connections (degree centrality) and betweenness centrality. It could be that more connected authors can leverage their social capital to diffuse their research and get more citations. Rotating leadership is also positively correlated with citations, supporting the idea that a bigger network dynamism of scholars is rewarded with more citations. Similarly, all network metrics related to the centrality of papers in the publication network significantly correlate with citations received. It could be that being highly cited is not just a matter of journal ranking, but also depends on the level of embeddedness in the two social networks we study. Consistently network constraint correlates negatively both for the author and the publication network, suggesting that when ego networks are more open, with more structural holes, there can be advantages of mediation across different social groups. Authors that have the power to link unconnected peers could be more effective in diffusing their ideas and research (Burt 2004). Similarly, papers that enable the collaboration of unconnected social groups could attract citations from a larger audience. On the other hand, metrics extracted from the analysis of paper abstracts seem to play a minor role; among them, abstract length is the one with the highest correlation. Of course these are just exploratory speculations, as correlation only reveals associations, without taking into account the combined effects of variables. For this reason, we extended the analysis with the intent of building a more comprehensive forecasting model that allows us the identification of future highly cited papers – in particular those which, six years after publication, receive a number of citations high enough to be in the uppermost quartile.

*Table 3: Spearman's correlation coefficients for variables used in the analysis*

| | Y | SJR | No. of authors | $X_1$ | $X_2$ | $X_3$ | $X_4$ | $X_5$ | $X_6$ | $X_7$ | $X_8$ | $X_9$ | $X_{10}$ | $X_{11}$ | $X_{12}$ | $X_{13}$ |
|---|---|---|---|---|---|---|---|---|---|---|---|---|---|---|---|---|
| Y | 1.000 | | | | | | | | | | | | | | | |
| SJR | .574** | 1.000 | | | | | | | | | | | | | | |
| No. of authors | .222** | .128** | 1.000 | | | | | | | | | | | | | |
| $X_1$ | .287** | .233** | .398** | 1.000 | | | | | | | | | | | | |
| $X_2$ | -.248** | -.207** | -.337** | -.835** | 1.000 | | | | | | | | | | | |
| $X_3$ | .108** | .108** | .148** | .367** | -.142** | 1.000 | | | | | | | | | | |
| $X_4$ | .253** | .206** | .462** | .582** | -.644** | .200** | 1.000 | | | | | | | | | |
| $X_5$ | .308** | .244** | .451** | .945** | -.822** | .315** | .564** | 1.000 | | | | | | | | |
| $X_6$ | -.305** | -.246** | -.517** | -.764** | .736** | -.271** | -.735** | -.820** | 1.000 | | | | | | | |
| $X_7$ | -.019** | -0.008 | .020** | -.102** | -.026** | .485** | -.078** | -.061** | .161** | 1.000 | | | | | | |
| $X_8$ | .303** | .250** | .320** | .859** | -.815** | .249** | .635** | .879** | -.782** | -.184** | 1.000 | | | | | |
| $X_9$ | .282** | .209** | .536** | .752** | -.603** | .268** | .615** | .737** | -.728** | -.126** | .660** | 1.000 | | | | |
| $X_{10}$ | .146** | .011** | .090** | -0.005 | 0.007 | -.042** | .030** | .016** | -.023** | -.011** | .017** | .019** | 1.000 | | | |
| $X_{11}$ | .050** | -.019** | .028** | .051** | -.045** | .010* | .029** | .048** | -.039** | -.011** | .048** | .046** | .252** | 1.000 | | |
| $X_{12}$ | 0.006 | -.084** | .040** | 0.004 | .012** | -.013** | -.019** | .013** | .020** | .016** | 0.003 | .008* | .591** | .134** | 1.000 | |
| $X_{13}$ | .031** | .098** | -.013** | 0.002 | -.021** | .020** | .031** | -0.001 | -.042** | -.009* | 0.004 | 0.006 | -.546** | -.120** | -.917** | 1.000 |
| $X_{14}$ | -.020** | -.012** | -.076** | .106** | -.080** | .010* | .038** | .077** | 0.005 | -.048** | .106** | .057** | -.042** | .109** | .051** | -.108** |

*Note.* Statistical significance: ***$p < 0.001$; **$p < 0.05$; *$p < 0.1$. Y = Citations, $X_1$ = Degree - publication network, $X_2$ = Constraint - publication network, $X_3$ = Closeness - publication network, $X_4$ = Betweenness - publication network, $X_5$ = Degree - author network, $X_6$ = Constraint - author network, $X_7$ = Closeness - author network, $X_8$ = Betweenness - author network, $X_9$ = Rotating Leadership, $X_{10}$ = Abstract Length, $X_{11}$ = Sentiment, $X_{12}$ = Complexity, $X_{13}$ = Diversity, $X_{14}$ = Commonness

We trained a parallel tree boosting machine learning model, namely XGBoost (Chen and Guestrin 2016), whose results are presented in Table 4. The model has been trained on 75% of observations and its performance has been subsequently evaluated considering the remaining 25% of data (out of sample). This process of random sampling without replacement of the training set and forecasting (on the remaining test set) has been repeated 300 times, i.e. we used Monte-Carlo cross validation (Dubitzky et al. 2007). We also evaluated the forecast performance of other algorithms, such as random forests (Breiman 2001), without getting to better results. Similarly, we tested other possible selections of highly cited papers – for example considering the upper quintile instead of quartile – and obtained results similar to those we present here.

Accuracy of predictions was quite good and stable across 300 random repetitions, with the model returning, on average, correct answers in 79.2% of cases, with an average score of 0.41 for the Cohen's Kappa and of 0.70 for the Area Under the ROC-curve. These results seem quite promising when compared with those reported by Abramo et al. (2019) on a dataset of publications submitted to the first Italian research assessment exercise (VTR 2006), exclusively based on peer review. Contrasting the peer review rating with long term citation scores, the authors obtained a 75% agreement and a Cohen's k equal to 0.172.

It is also important to notice that our main goal was not to obtain a 100% accurate model; more than finding the perfect forecast, we were interested in identifying variables that could be more relevant when predicting citations. Accordingly, Table 4 shows the importance of each predictor, calculated as the average of its absolute SHAP values (S. M. Lundberg and Lee 2017): the higher the score reported in the table, the more important the predictor. SHAP stands for SHapley Additive exPlanations and is a well-known evaluation approach, applicable to the output of different machine learning models. This method showed better consistency than previous approaches (S. M. Lundberg and Lee 2017) and proved to be particularly appropriate for tree ensembles (S. M. Lundberg et al. 2020). These last analyses were carried out using the Python programming language, specifically the packages SHAP (S. M. Lundberg and Lee 2017) and XGboost (Chen and Guestrin 2016).

Consistent with the results of the correlation analysis, we find that journal ranking is the most important predictor of highly cited papers, followed by rotating leadership, the number of authors and betweenness centrality in the publication network. It seems that social capital plays a role in terms of authors' direct connections with peers, who could read and cite their papers. Keeping a dynamic position is also important. In addition, papers which result from the collaboration of different social groups also get more citations. Lastly, writing longer, more informative abstracts seems to contribute a little to the improvement of model performance. The other variables, on the other hand, contribute little to our model.

*Table 4: Feature importances*

| Variable | Mean SHAP Values | SD SHAP Values |
| --- | --- | --- |
| SJR | 1.421 | 0.025 |
| No. of Authors | 0.626 | 0.049 |
| $X_1$, Degree - publication network | 0.070 | 0.018 |
| $X_2$, Constraint - publication network | 0.053 | 0.012 |
| $X_3$, Closeness - publication network | 0.047 | 0.011 |
| $X_4$, Betweenness - publication network | 0.309 | 0.028 |
| $X_5$, Degree - author network | 0.042 | 0.009 |
| $X_6$, Constraint - author network | 0.047 | 0.009 |
| $X_7$, Closeness - author network | 0.064 | 0.013 |
| $X_8$, Betweenness - author network | 0.068 | 0.012 |
| $X_9$, Rotating Leadership | 0.808 | 0.048 |
| $X_{10}$, Abstract length | 0.169 | 0.012 |
| $X_{11}$, Sentiment | 0.079 | 0.010 |
| $X_{12}$, Complexity | 0.097 | 0.013 |
| $X_{13}$, Diversity | 0.103 | 0.014 |
| $X_{14}$, Commonness | 0.063 | 0.007 |

Journal ranking is by far the most important feature to forecast future citations and scholarly impact. Indeed, we notice that our sample comprises about 8,000 papers which are both highly cited and published in top journals. However, a smaller number of papers, about 500, has the peculiar characteristic of being highly cited even if published in journals that have very low rankings (bottom 25% of the SJR distribution). How is that possible? We explored the differences between these two sets of papers through the t-tests presented in Figure 1.

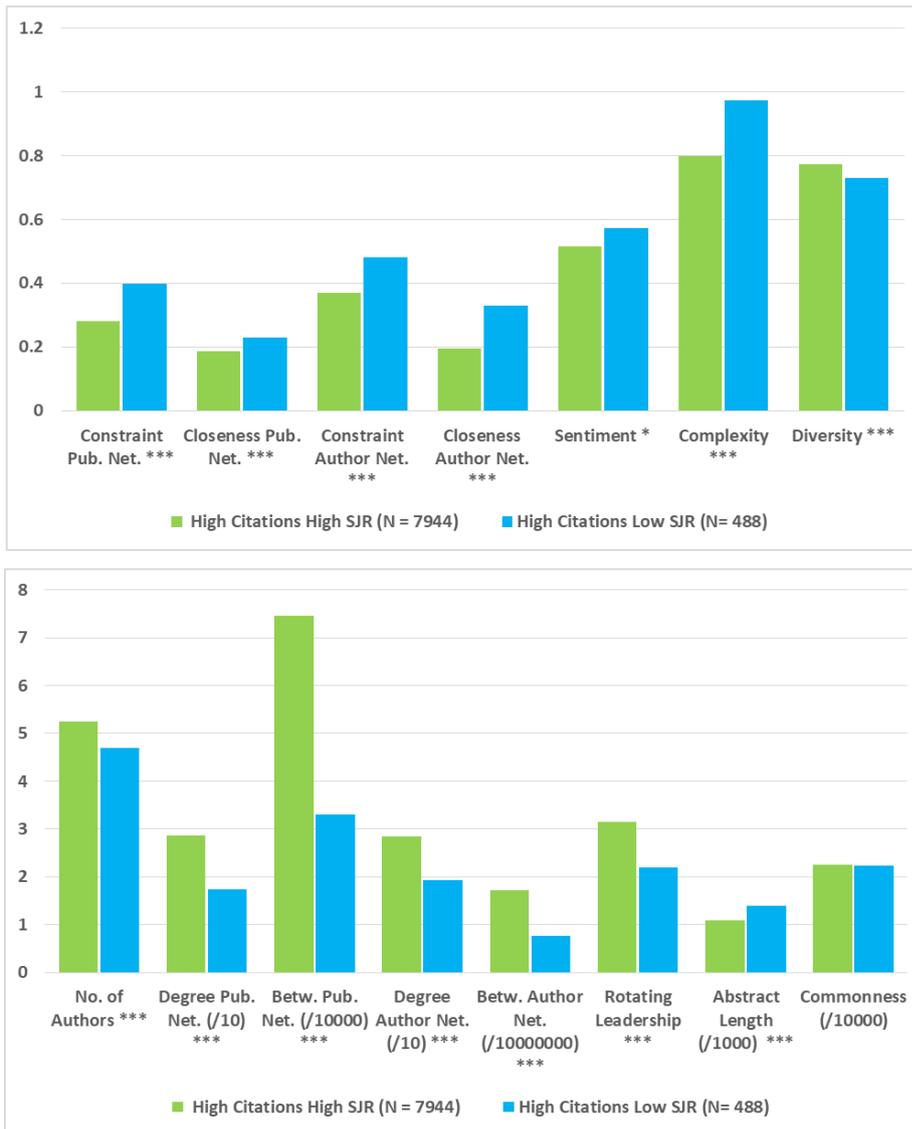

*Figure 1. Characteristics of highly cited papers published in low ranked journals*
*(T-tests, \*\*\*p < .001; \*p < .05).*

Apart from commonness, all the variables are significantly different. Successful papers published in low SJR journals seem to present more positive results (higher sentiment in the abstract) and new ideas (higher complexity), and have longer abstracts (even if this could be influenced by journal policies). Both these papers and their authors are closer to the network core (closeness is higher). Surprisingly, the number of authors and their connections – as well as betweenness centrality and rotating leadership – are lower with respect to papers in the top citations quartile, published in top journals. It seems that focused network embeddedness is the major driver of success for this set of papers (low SJR, high citations). It is not just a matter of being close to the network core, but also being part of a compact group with few structural holes. We speculate that in these cases unity is strength.

## 5. Discussion and Conclusions

In our research, we examined several characteristics of scientific papers which help predict their scholarly impact six years after publication. Results of our parallel tree-boosting machine learning model confirm findings of previous research, which indicate that journal impact factor and number of authors have a significant and positive effect on citations (Abramo and D'Angelo 2015; Bornmann et al. 2014; Leimu and Koricheva 2005; Mingers and Xu 2010; Waltman and van Eck 2015). We used these metrics as control variables and combined them with measures of social network and semantic analysis, which allowed the identification of highly cited papers with 79.2% accuracy. We found that authors' social capital has a role in attracting citations, thus publishing papers with well-connected authors can be an advantage. However, this effect is relatively small if compared with authors' rotating leadership, i.e. the ability to frequently change position in the collaboration network, moving back and forth from center to periphery. Indeed, authors' rotating leadership (change in betweenness centrality) emerged as one of the most important predictors of highly cited papers: it is not just a matter of authors' brokerage power, i.e. the ability to bridge connections across different social groups; authors' ability to activate bridging collaborations and subsequently leave space to others, without keeping dominant or static positions, was the third most important predictor. This is consistent with previous research showing that rotating leaders foster community growth and participation (Antonacci et al. 2017) and that dynamic social styles can favor innovation and knowledge sharing (Allen et al. 2016; Davis and Eisenhardt 2011). Accordingly, our study extends the research on the forecasting of scholarly impact, giving evidence to the contribution of new metrics of social network analysis, such as rotating leadership. In particular, we analyzed two social networks over a period of three years: the first, linking authors based on their scientific collaborations; the second, considering the social position of scientific papers based on their shared authors. The analysis of this second network revealed another important factor of publication success: scientific papers that resulted from the collaboration of different social groups – whose betweenness centrality was therefore higher – were more frequently ranked among the highly cited papers.

Predictors related to the semantic analysis of paper abstracts exhibited a lower, yet significant, importance. In particular, longer and more informative abstracts, whose texts have a higher lexical diversity, seem to attract more citations. In this regard, our findings are aligned with research showing that shorter abstracts lead to fewer citations (Weinberger et al. 2015) and contrast with the study of Letchford et al. (2016) which proves the opposite. Our results also support the idea that abstracts that are more creative and diversified can attract more citations, as discussed by Freeling and colleagues (2019).

As a last step of analysis, we examined those papers which represented an exception to the idea that journal ranking plays a major role in attracting citations. In particular, we found about 500 articles which were published in low SJR journals but were highly cited. We compared them with highly cited papers published in top journals. Distinctive characteristics of successful low-SJR papers are that they present more positive results – abstract sentiment is higher on average – and have longer and more complex abstract texts, thus probably being even more informative than regular highly cited papers. Authors of these papers are close to the network core (high closeness); however, their rotating leadership is surprisingly lower than the one of authors of highly cited papers published in top journals. These publications also rarely involve scholars of different social groups. It seems that successful papers published in low-ranked journals mostly benefit from focused network embeddedness of their authors. Being part of a closed group with few structural holes, and being close to the network core, seem much more important than bridging social ties.

Our work not only extends research on the forecasting of paper citations, but also contributes to the identification of new metrics derived from social network and semantic analysis. The study has several limitations and the results of our analysis do only give limited insights about causality – which should be examined in future research. Is it that well-connected authors will get more citations in the future – one would assume this is true, or is it that highly cited papers will lead to more centrality for authors – one would assume that this is also true.

Compared to past studies (Abramo, D'Angelo, and Felici 2019; Bornmann et al. 2014; Bruns and Stern 2016; Levitt and Thelwall 2011; Stegehuis et al. 2015; D. Wang et al. 2013), we present a model that considers the combined effects of a high number of predictors, i.e. scientific paper features. Future research could use our model and predictors to examine citations dynamics in fields other than chemical engineering, or consider even more control variables, to account, for example, for the presence of sleeping beauties or for possible geographical biases (Wuestman et al. 2019). Working with different citation timeframes, could reveal new factors impacting paper success. Subcategories of articles could also be considered, distinguishing between research papers and reviews of literature (we have already excluded the other categories of documents). Moreover, it might be that open access papers are cited more, as they are more easily accessible than paywalled ones (Eysenbach 2006) – even if, nowadays, this effect is mitigated by many factors, such as the increased availability of pre-print versions of published papers[6] and the existence of (pirate) websites like Sci-Hub (Himmelstein et al. 2018). To the extent of our knowledge, this is one of the first studies where sentiment analysis of scientific abstracts is carried out. Indeed, sentiment analysis of scientific papers is a new and interesting problem (Athar 2011; Athar and Teufel 2012). Scientific communication is usually fact-based, and more technical, than texts that can be mined from other sources (Athar 2011) – for example social media. In this sense, our research is partially exploratory and tries to see whether the sentiment metric conveys any useful information for the prediction of future citations. We calculated sentiment using the VADER lexicon (Hutto and Gilbert 2014), whereas future research could consider different, or context-specific, approaches.

---

[6] https://arxiv.org/stats/monthly_submissions